\documentclass{article}

\usepackage[preprint]{neurips_2026}

\usepackage[utf8]{inputenc}
\usepackage[T1]{fontenc}
\usepackage{amsmath,amssymb}
\usepackage{booktabs}
\usepackage{graphicx}
\usepackage{xcolor}
\usepackage{url}
\usepackage{hyperref}

\newcommand{\seq}{x}
\newcommand{\struct}{s}
\newcommand{\condition}{c}
\newcommand{\maskedstruct}{\tilde{s}}
\newcommand{\maskedseq}{\tilde{x}}
\newcommand{\model}{p_\theta}
\newcommand{\prior}{p_0}
\newcommand{\oracle}{q}
\newcommand{\given}{\,\vert\,}
\newcommand{\nt}{\mathrm{nt}}

\title{GoForth: Language Models for RNA Design under Structure, Sequence, and Coding Constraints}

\author{%
  Michael Lindsey \\
  UC Berkeley and LBNL \\
  Berkeley, CA \\
  \texttt{lindsey@berkeley.edu} \\
}

\begin{document}

\maketitle

\begin{abstract}
RNA inverse sequence design has broad biological and engineering applications, but computational methods for practical design queries remain limited. Such queries may impose several constraints at once, including target folds or motifs, fixed bases, and coding restrictions, while leaving arbitrary sequence and structure in unspecified regions. Because these constraints may permit many acceptable sequences, we study RNA design as a conditional generative modeling problem. The basic object is a conditional law over RNA sequences given a user-specified condition, with full inverse folding as a special case. We introduce GoForth, a forward-trained RNA language model that conditions on structure, sequence, and coding targets. The formulation separates three ingredients that are often entangled in RNA design: a sequence prior, a forward folding sampler, and a reward or likelihood oracle. We train encoder-decoder models on witnessed folds rather than on outputs from an inverse-design teacher and validate our methodology on full inverse-folding benchmarks, as well as tasks involving constraints on structure, sequence, and coding. The resulting models achieve fast and high-quality candidate generation for mixed RNA design specifications. Moreover they furnish useful semantic embeddings of design tasks and a robust learned notion of designability.
\end{abstract}

\section{Introduction}

RNA design seeks to choose a sequence of nucleotides such that the resulting RNA molecule satisfies structural or functional constraints. At least for pseudoknot-free secondary structure, the forward thermodynamic problem is tractable under structured energy models: packages such as ViennaRNA and NUPACK compute minimum-free-energy structures, partition functions, base-pair probabilities, and constrained ensembles by dynamic programming \citep{lorenz2011viennarna,zadeh2011nupack,zadeh2011ensemble}. The inverse problem is different. A target may have many compatible sequences, no compatible sequence under a chosen success criterion, or many qualitatively different ways to be specified.

Full inverse folding is the cleanest benchmark: given a complete secondary structure, find a sequence whose predicted fold matches it. This setting has driven classical search methods, reinforcement-learning methods, Eterna-derived benchmarks, and recent language-model approaches \citep{eastman2018rl,runge2019learna,koodli2019eternabrain,gautam2026designing}. It is not, however, the only useful interface. Designers often specify local motifs, accessible sites, protected windows, fixed bases, or coding constraints while leaving other degrees of freedom open. Existing tools cover important slices of this space, though they are typically based on combinatorial search, and there is no simple design protocol for arbitrary mixed structure, base, and coding constraints.

We introduce GoForth, trained simply on sequence-condition pairs $(\seq,\condition)$ derived by masking a witnessed fold. The source of these pairs can be synthetic or experimental folds, which furnish some condition $c$ given $x$ drawn from a prior distribution. Our forward-trained objective {does not} rely on the evaluation of the thermodynamic or likelihood oracle \(q(\condition\given\seq)\), scoring whether a candidate sequence realizes a condition. It moreover does not rely on any inverse-design teacher, which would have to return a good sequence for any given condition. This separation is especially important for arbitrary partial and mixed constraints, where there may be no strong teacher algorithm analogous to a full-structure inverse-folding solver.

\paragraph{Contributions.}
We make five contributions.
First, we formulate RNA design under structure, sequence, and protein-coding constraints as conditional sequence generation under a compositional condition language, with full inverse folding as one slice rather than the defining case.
Second, we train encoder-decoder proposal models from forward/witnessed sequence-condition pairs, avoiding the need for a specialized inverse-design teacher for each constraint family.
Third, we calibrate the models on full inverse folding against RNA-Design-LM, reporting thermodynamic quality, parameter counts, and neural generation runtime.
Fourth, we advance new witnessed benchmarks for base- and coding-aware design, so failures can be interpreted against feasible specifications. We comment that, to the best of our knowledge, no standard benchmarks or algorithms have been widely accepted for partial design tasks involving base and coding constraints, and in particular amortized approaches have not been considered.
Fifth, we show that learned condition encodings contain meaningful semantic information and a transferable ``designability signal.''

Code and datasets are maintained at \texttt{\href{https://github.com/quantumtative/GoForth}{github.com/quantumtative/GoForth}}.

\section{Related Work}

\paragraph{Thermodynamic RNA design.}
Classical RNA inverse-design methods combine folding or partition-function oracles with per-target optimization. RNAinverse, RNA-SSD, INFO-RNA, RNAiFold, NUPACK design, and antaRNA represent dynamic-programming, local-search, constraint-programming, and stochastic-search approaches to the problem \citep{andronescu2004rnassd,busch2006inforna,garciamartin2013rnaifold,dotu2014rnaifold,backofen2015antarna}. NUPACK also supports constrained multistate sequence design for nucleic-acid reaction pathways, a different but important formulation of physically constrained design \citep{wolfe2017constrained}. More recent work emphasizes ensemble objectives and designability diagnostics. SAMFEO optimizes target probability or ensemble defect and is a major component in the RNA-Design-LM training pipeline \citep{zhou2023samfeo,gautam2026designing}. SamplingDesign optimizes a distribution over valid sequences for a full target structure, but remains a per-target optimization method rather than an amortized generator for arbitrary mixed conditions \citep{tang2026samplingdesign}. Fitness-function, undesignability, and probabilistic designability studies further show why minimum free energy (MFE) structure matching alone is an incomplete objective \citep{ward2023fitness,zhou2024undesignable,zhou2025minimalundesignable,zhou2026probdesignability}.

\paragraph{Partial and constrained design.}
Partial RNA design is not new. libLEARNA formulates RNA sequence and structure motifs with wildcards, includes test-time adaptation, and defines a heterogeneous suite of partial-design tasks \citep{runge2024partial}. RNAblueprint and RNARedPrint sample sequences under structural and sequence constraints, including multi-state specifications \citep{hammer2017rnablueprint,hammer2019rnaredprint}, while NUPACK emphasizes thermodynamic multistate design \citep{wolfe2017constrained}. Coding constraints are usually treated in the mRNA-design literature, where objectives include minimum free energy (MFE) structure, ensemble free energy, codon usage, or expression-related proxies \citep{zhang2023lineardesign,dai2025ensembledesign,fornace2026direct}. However, to the best of our knowledge, a general amortized approach for partial design has not been formulated, perhaps due to the lack of consensus around teacher methods, which typically rely on search heuristics. Indeed a recent benchmark~\citep{cole2024rnainvbench} for full structure design leaves benchmarking for base-constrained structure design open.

\paragraph{Language models.}
Gautam et al. recently showed that RNA inverse folding can be framed as conditional language modeling with constrained decoding and reinforcement learning \citep{gautam2026designing}. Remarkably, their RNA-Design-LM system uses a decoder-only language model initialized from a pretrained \emph{natural-language} model and then adapted to full-structure inverse folding, with supervised data from a strong inverse-design teacher and a subsequent oracle-based RL stage. Relative to their work, GoForth is trained from scratch as an encoder-decoder model on witnessed \((\seq,\struct)\) or \((\seq,\condition)\) pairs. This architecture separates condition encoding from autoregressive sequence generation, supports masked structure/base/coding channels, and gives encoder embeddings that can be reused for designability analysis. 

Meanwhile, RNA foundation models and 3D inverse-design models could potentially define complementary sources of priors and folding oracles, but they do not address the inverse RNA design problems considered here \citep{chen2022rnafm,penic2025rinalmo,tan2024rdesign,joshi2025grnade}.



\section{Problem Formulation}

Let \(\seq=(x_1,\ldots,x_n)\) be an RNA sequence over \(\{\texttt{A},\texttt{C},\texttt{G},\texttt{U}\}\). Let \(\struct\) be a pseudoknot-free secondary structure, represented either as dot-bracket notation or as a side-token string over left-paired, right-paired, and unpaired sites. For a fixed \(\seq\), a thermodynamic model defines an energy \(E(\struct;\seq)\), a partition function \(Z(\seq)\), and a Boltzmann law
\begin{equation}
  \oracle(\struct\given\seq)
  =
  Z(\seq)^{-1}\exp[-E(\struct;\seq)/(kT)] .
\end{equation}
All thermodynamic claims in this paper are relative to the chosen oracle. In the experiments below, the oracle is ViennaRNA 2.

\paragraph{Bayesian inverse design.}
For a complete target structure $s$, a natural design distribution is the posterior 
\begin{equation}
  \pi(\seq\given\struct)
  \propto
  \oracle(\struct\given\seq) \, \prior(\seq),
  \label{eq:full-posterior}
\end{equation}
where \(\prior\) is a sequence prior. MAP design maximizes \(\log \oracle(\struct\given\seq)+\log \prior(\seq)\); sampling instead exposes a family of candidates that balance prior likelihood with likelihood of the observation $s$. Typical inverse design approaches can be viewed in this sense as MAP design with a flat prior. Indeed, throughout we use the flat prior, i.e., a uniform i.i.d. prior over sequence nucleotides. This is a neutral baseline that stands in roughly for biological plausibility in that base counts tend to be balanced and long repeat sequences tend not to appear. Replacing it with a learned natural-RNA prior defines an interesting extension, especially when weak conditions leave many degrees of freedom.

For a general condition \(\condition\), possibly involving full or partial constraints on the structure $s$ or sequence $x$, our design distribution is 
\begin{equation}
  \pi(\seq\given\condition)
  \propto
  \oracle(\condition\given\seq) \, \prior(\seq).
  \label{eq:condition-posterior}
\end{equation}
Here
\begin{equation}
  \oracle(\condition\given\seq)
  :=
  \sum_{ \struct:\,(x,s)\mapsto\condition}
  \oracle(\struct\given\seq)
\end{equation}
is the marginal distribution of the constraint given the sequence, and \( (x, \struct) \mapsto\condition\) means that \((x,\struct)\) satisfies all specified structure, sequence, and coding constraints in \(\condition\). 
For the constraints considered in this work, this $q(c \, \vert \,x )$ can be computed, and conditional structures can be sampled, by constrained dynamic programming in ViennaRNA.

\paragraph{Condition languages.}
GoForth uses a sequence of condition tokens. The structure  channel $\tilde{s}$ consists of \(\texttt{L}\), \(\texttt{R}\), \(\texttt{x}\), \(\texttt{\#}\), and \(\texttt{?}\) tokens, denoting left-paired, right-paired, unpaired, paired-with-unknown-orientation, and unconstrained/masked positions. The base channel $\tilde{x}$ consists of \(\texttt{A},\texttt{C},\texttt{G},\texttt{U},\texttt{?}\). The coding-aware model additionally receives, at each nucleotide position, a frame token \(\rho_i\in\{0,1,2\}\) and an amino-acid token \(\psi_i \in \{ \textrm{Ala}, \textrm{Arg}, \ldots, \textrm{Val}, \textrm{Stop}, \texttt{?}  \}\) giving the required residue for the codon containing that nucleotide, or a null/unknown token outside revealed coding constraints. We focus on the following model families:
\begin{align}
  \text{GoForth-FS} &: \model(\seq\given\struct),\\
  \text{GoForth-PSB} &: \model(\seq\given\maskedstruct,\maskedseq),\\
  \text{GoForth-PSBC} &: \model(\seq\given\maskedstruct,\maskedseq,\rho,\psi),
\end{align}
where \(\rho\) encodes reading frame and \(\psi\) encodes revealed amino-acid constraints. Here FS denotes `full structure constraints,' PSB `partial structure and base constraints,' and PSBC `partial structure, base, and coding constraints.' Alternative model classes are mentioned in Appendix~\ref{app:model-details}.

\paragraph{Inference and post-training.}
At inference time we draw \(N\) autoregressive samples, optionally using hard masks that enforce fixed bases, valid base-pair compatibility, and codon constraints. A sampling temperature \(\tau\) rescales token logits before sampling. This is a heuristic but useful knob: low \(\tau\) concentrates the learned proposal, while higher \(\tau\) gives diversity.

When a reward \(R(\seq;\condition)\) is available, post-training can be achieved through the KL-regularized objective
\begin{equation}
  \max_\pi
  \mathbb{E}_{\seq\sim\pi}[R(\seq;\condition)]
  - \lambda_{\mathrm{KL}}\,
  \mathrm{KL}\!\left(
    \pi(\cdot\given\condition)\middle\|\pi_0(\cdot\given\condition)
  \right),
  \label{eq:kl-objective}
\end{equation}
whose exact optimum is proportional to
\begin{equation}
  \pi_0(\seq\given\condition) \, \exp(R(\seq;\condition)/\lambda_{\mathrm{KL}}).
\end{equation}
In our approach, post-training does not offer any advantage relative to tuning the temperature $\tau$. Instead we conduct post-training experiments using the reward $R(x;c) = \log q(c\,\vert\,x)$ as a way to \emph{validate} that the temperature parameter $\tau$ successfully anneals from the posterior to an approximate MAP estimator. To demonstrate this, we use group relative policy optimization (GRPO) as an approximate policy-optimization method for this tilt \citep{shao2024deepseekmath}. The coefficient \(\lambda_{\mathrm{KL}}\) is temperature-like: smaller values sharpen the oracle tilt away from the pretrained proposal. For our choice of $R$, the exact tilt is \(\pi_0(\seq\given\condition) \, \oracle(\condition\given\seq)^{1/\lambda_{\mathrm{KL}}}\), so GRPO acts like posterior annealing when \(\pi_0\) already approximates \(\pi(\seq\given\condition)\). In these experiments, post-training targets are sampled from synthetic reservoirs rather than from named benchmark instances. Robust convergence is validated in Appendix~\ref{app:model-details}.

\section{Implementation}

\paragraph{Architecture.}
All models are Transformer encoder-decoder autoregressive sequence models \citep{vaswani2017attention}; Appendix~\ref{app:model-details} gives the model architecture and training policy details. The encoder reads the condition string and the decoder generates RNA bases left to right. The small models use 8 layers and width 512, with about 59M parameters for the full/partial-structure models and 60M for the coding-aware model. The large models use 12 layers and width 768, with about 200M parameters. Training uses conditional likelihood
\begin{equation}
  \log \model(\seq\given\condition)
  =
  \sum_{i=1}^L \log p_\theta(x_i\given x_{<i},\condition),
\end{equation}
which is maximized in aggregate over a training data reservoir.

\paragraph{Training data.}
The present experiments use unpseudoknotted synthetic structure-sequence pairs $(x^{(n)}, s^{(n)})$ generated from the ViennaRNA forward distribution, conditioned on $x^{(n)}$ drawn from the uniform prior with uniformly random lengths between  \(11\)-\(310\, \nt\). Longer examples are used only for length-generalization evaluation. Training sees only paired examples $(x^{(n)}, c^{(n)})$, not benchmark targets or sequences produced by an inverse-folding teacher, where partial conditions $c^{(n)}$ are obtained by randomly masking structure, base, and coding channels of a witness sequence, according to the model category; Appendix~\ref{app:benchmark-details} gives the masking policies. GRPO post-training uses random synthetic targets from the same reservoirs, not from benchmark datasets.

\paragraph{Evaluation.}
For each target and method, we generate a batch of \(N\) candidates, rescore them with ViennaRNA, and report best-of-\(N\) quality. The main tables use \(N=100\); metric definitions and benchmark composition are detailed in Appendices~\ref{app:model-details} and \ref{app:benchmark-details}. Pretrained experiments use \(\tau=0.1\) unless stated otherwise, while GRPO experiments use the temperature at which they were post-trained.

\section{Experiments}

\subsection{Full-Structure Inverse Folding}

Full inverse folding gives a standardized calibration point for GoForth against specialized methods. Table~\ref{tab:full-benchmark} reports the 1031-target training-length aggregate; Table~\ref{tab:length-generalization} in Appendix~\ref{app:length} separates targets longer than the \(310\,\nt\) training range.

\begin{table}[t]
  \centering
  \small
  \caption{Full-structure calibration benchmark at \(N=100\) samples. Entries are means over the 1031 targets with length at most \(310\,\nt\); timing excludes Vienna rescoring.}
  \label{tab:full-benchmark}
  \begin{tabular}{lllrrrr}
    \toprule
    Method & Params & Scope & gen. sec/target & MFE hit & NED & Best \(q(\struct\given\seq)\) \\
    \midrule
    RNA-Design-LM SL & 500M & FS & 1.46 & 0.777 & 0.0317 & 0.665 \\
    RNA-Design-LM SL+RL & 500M & FS & 1.38 & \textbf{0.805} & \textbf{0.0175} & \textbf{0.722} \\
    GoForth-FS small $\tau = 0.1$ & 59M & FS & 0.25 & 0.775 & 0.0251 & 0.621 \\
    GoForth-FS small GRPO & 59M & FS & 0.26 & 0.761 & 0.0213 & 0.687 \\
    GoForth-FS large $\tau = 0.1$ & 200M & FS & 0.41 & 0.772 & 0.0273 & 0.596 \\
    GoForth-PSB small $\tau = 0.1$ & 59M & PSB & 0.25 & 0.742 & 0.0251 & 0.615 \\
    GoForth-PSB large $\tau = 0.1$ & 200M & PSB & 0.41 & 0.758 & 0.0236 & 0.629 \\
    GoForth-PSB large GRPO & 200M & PSB & 0.41 & 0.741 & 0.0247 & 0.624 \\
    GoForth-PSBC small $\tau = 0.1$ & 60M & PSBC & \textbf{0.24} & 0.749 & 0.0243 & 0.628 \\
    GoForth-PSBC large $\tau = 0.1$ & 201M & PSBC & 0.41 & 0.749 & 0.0242 & 0.624 \\
    GoForth-PSBC large GRPO & 201M & PSBC & 0.40 & 0.731 & 0.0255 & 0.611 \\
    \bottomrule
  \end{tabular}
\end{table}

RNA-Design-LM SL+RL is strongest on this full-structure benchmark. However, we comment that RNA-Design-LM SL+RL was post-trained on a curated dataset. When compared to RNA-Design-LM, which trains only on the same type of procedurally generated data as our methods, we perform similarly (typically with better NED and worse likelihood) at significantly greater speed. We want to avoid curating a dataset because benchmarks are too open-ended for partial structure design tasks and moreover because there is some risk of test set contamination, though the authors \citep{gautam2026designing} screen out obvious sources. We point to the semantic embedding discussed in Section~\ref{sec:embedding} as a potential avenue for exploring dataset curation in future work.

Both RNA-Design-LM methods have access to an inverse teacher that may not be available for partial design tasks. Our comparable performance on full structure design given only weak inputs suggests that similar success can be extended to more general tasks.

GRPO gives no uniform improvement; with rewards based on \(q(\condition\given\seq)\), this is consistent with low-temperature best-of-\(N\) sampling already approximating the useful annealed posterior.

See Appendix~\ref{app:length} for a discussion of generalization to larger lengths, inference time scaling, and performance on a slice of natural data.

\subsection{Partial-Design Benchmarks: Structure, Sequence, and Coding}

In the absence of standard benchmarks, we put forward a slate of witnessed partial RNA design tasks. These are constructed by computing $(x,s)$ in which $s$ is the Vienna MFE structure for $x$, then masking $(x,s)$ to produce a condition $c$. We consider several masking policies outlined in Appendix~\ref{app:benchmark-details}. In Table~\ref{tab:partial-procedural}, we consider the same artificial masking policies that we used in training, though with different data. In Table~\ref{tab:partial-bprna-annotation}, we consider applying this policy to the \emph{actual} experimental $(x,s)$ measured in the bpRNA dataset, defining an experiment that is \emph{not} Vienna-witnessed and accordingly more difficult to fit within the Vienna model.

Meanwhile, we also consider masking policies based on biologically legible motifs such as terminal hairpins, fixed loop bases, pairedness halos, and codon-window structure under peptide preservation, reporting results in Table~\ref{tab:partial-structured}. Appendix~\ref{app:benchmark-details} gives the construction details. Among these masking policies, Policy 1 is nearly solved; Policy 2 is a more discriminating structure/base task; Policy 3 demonstrates the coding-aware use case.

\begin{table}[t]
  \centering
  \small
  \caption{Artificially masked, Vienna-witnessed partial-design benchmarks, \(N=100\). `MFE hit' measures whether any candidate satisfies all \emph{active}/\emph{unmasked} structure tokens and hard base/codon constraints. `UMFE hit' measures when this holds and the MFE is unique.}
  \label{tab:partial-procedural}
  \resizebox{\textwidth}{!}{%
  \begin{tabular}{llrrrrrr}
    \toprule
    Benchmark & Method & MFE hit & UMFE hit & Best \(q(c\given x)\) & Best MFE error & Candidate hit frac. & gen. sec/target \\
    \midrule
    bpRNA & GoForth-PSB large $\tau = 0.1$ & 0.70 & 0.56 & 0.373 & 0.010 & 0.370 & 1.73 \\
    bpRNA & GoForth-PSBC large $\tau = 0.1$  & {0.72} & 0.50 & 0.282 & 0.014 & 0.383 & 2.13 \\
    Coding & GoForth-PSBC large $\tau = 0.1$  & 0.81 & 0.62 & 0.476 & 0.006 & {0.499} & 1.57 \\
    \bottomrule
  \end{tabular}}
\end{table}

\begin{table}[t]
  \centering
  \small
  \caption{Partial design benchmark for artificially masked bpRNA data with \emph{experimental} annotation, \(N=100\). Note that the data is not Vienna-witnessed, hence more likely to be undesignable. Success is ViennaRNA MFE satisfaction of active annotation tokens.}
  \label{tab:partial-bprna-annotation}
  \resizebox{\textwidth}{!}{%
  \begin{tabular}{llrrrrrr}
    \toprule
    Benchmark & Method & MFE hit & UMFE hit & Best \(q(c\given x)\) & Best MFE error & Candidate hit frac. & gen. sec/target \\
    \midrule
    bpRNA-Annotation & GoForth-PSB large, \(\tau=0.1\) & 0.45 & 0.40 & 0.317 & 0.080 & 0.222 & 1.77 \\
    bpRNA-Annotation & GoForth-PSBC large, \(\tau=0.1\) & 0.27 & 0.22 & 0.177 & 0.119 & 0.131 & 2.18 \\
    \bottomrule
  \end{tabular}}
\end{table}

\begin{table}[t]
  \centering
  \small
  \caption{Motif-masked partial-design benchmarks, \(N=100\). P1 reveals terminal hairpins, P2 adds loop bases and a pairedness/accessibility halo, and P3 composes a codon-window structure with full peptide preservation.}
  \label{tab:partial-structured}
  \resizebox{\textwidth}{!}{%
  \begin{tabular}{lllrrrrr}
    \toprule
    Policy & Dataset/masking & Method & Targets & MFE hit & UMFE hit & Best \(q(c\given x)\) & Best MFE error \\
    \midrule
    P1 & bpRNA hairpins & GoForth-PSB large, \(\tau=0.1\) & 76 & 1.000 & 1.000 & 0.991 & 0.000 \\
    P1 & bpRNA hairpins & GoForth-PSBC large, \(\tau=0.1\) & 76 & 1.000 & 1.000 & 0.992 & 0.000 \\
    P2 & bpRNA loop+halo & GoForth-PSB large, \(\tau=0.1\) & 76 & 0.803 & 0.803 & 0.747 & 0.015 \\
    P2 & bpRNA loop+halo & GoForth-PSBC large, \(\tau=0.1\) & 76 & 0.816 & 0.789 & 0.731 & 0.016 \\
    P3 & Coding-window & GoForth-PSBC large, \(\tau=0.1\) & 100 & 0.800 & 0.680 & 0.557 & 0.010 \\
    \bottomrule
  \end{tabular}}
\end{table}

\begin{table}[h]
  \centering
  \small
  \caption{libLEARNA comparison on the lifted bpRNA-Annotation structure/base benchmark, \(N\le100\). The lifted targets contain no \(\texttt{\#}\) tokens. libLEARNA candidates are the top 100 emitted sequences per target under its own reward within a 20-second per-target optimizer budget (though this is exceeded in practice), then rescored with the ViennaRNA partial-condition scorer.}
  \label{tab:liblearna-lifted}
  \resizebox{\textwidth}{!}{%
  \begin{tabular}{lrrrrrr}
    \toprule
    Method & Targets & MFE Hit & UMFE Hit & Best \(q(c\given x)\) & Best MFE error & Wall sec/target \\
    \midrule
    libLEARNA PPO, 20s timeout & 100 & 0.140 & 0.090 & 0.0735 & 0.275 & 92.7 \\
    GoForth-PSB large, \(\tau=0.1\) & 100 & \textbf{0.400} & \textbf{0.350} & \textbf{0.274} & \textbf{0.0867} & 2.12 \\
    \bottomrule
  \end{tabular}}
\end{table}

To compare our method against the existing landscape, we also converted the bpRNA-Annotation PSB benchmark to a libLEARNA-compatible form by replacing paired-unknown \(\texttt{\#}\) tokens with their known left/right orientation. Note moreover that libLEARNA can handle base but not coding constraints.

As reported in Table~\ref{tab:liblearna-lifted}, GoForth significantly outperforms libLEARNA both in terms of speed and accuracy. The libLEARNA wall time is larger than the nominal optimizer timeout because each target launches a fresh optimization process with substantial setup and restart overhead.

The benchmarks introduced in this section are not direct leaderboards against a mature standard. Strong full-structure methods such as SAMFEO are not applicable to these partially specified tasks. libLEARNA is arguably the closest analogue for partial structure/base design, but its condition language does not cover coding constraints and does not use the paired-unknown \(\texttt{\#}\) token used by GoForth.

\subsection{Illustration of Annealing and Design Hacking}

\begin{figure}
  \centering
  \begin{minipage}{0.51\textwidth}
    \centering
    \includegraphics[width=\linewidth,trim={0 0mm 0 0},clip]{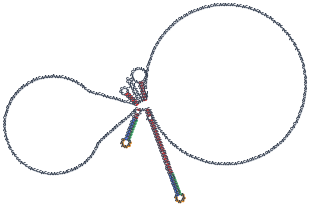}
    \small \(\tau=0.01\) MFE
  \end{minipage}
  \hfill
  \begin{minipage}{0.31\textwidth}
    \centering
    \includegraphics[width=\linewidth,trim={0 0mm 0 0},clip]{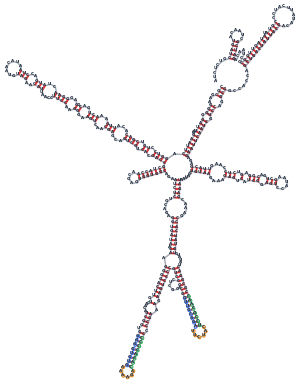}
    \small \(\tau=0.1\) MFE
  \end{minipage}
  \hfill
  \begin{minipage}{0.11\textwidth}
    \centering
    \includegraphics[width=\linewidth]{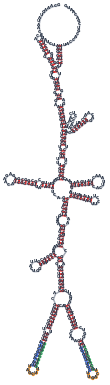}
    \small \(\tau=1.0\) MFE
  \end{minipage}
  \caption{Best-of-\(100\) ViennaRNA MFE layouts for the \(L=450\) two-hairpin demo. The specified structure is small relative to the sequence length. Blue and green denote constraints for left- and right-pairing, while yellow denotes unpaired constraint.}
  \label{fig:hairpin-svgs}
\end{figure}

In Figure \ref{fig:hairpin-svgs} we show how the sampling temperature $\tau$ influences GoForth's design selection for a highly underspecified task requesting only two hairpins. Note also from the figure that this simple task is an example of length generalization beyond the training window. Anecdotally, we report that very sparse constraint motifs generalize quite easily.

\subsection{Embedding Geometry, Designability, and Adversarial Prompting} \label{sec:embedding}

For a condition $c$, we mean-pool the encoder states \(h_1 (c),\ldots,h_n (c)\), subtract a training-block mean, and normalize to unit length. The resulting embeddings support dataset similarity metrics, nearest-neighbor inspection, and difficulty prediction.

\begin{figure}
  \centering
  \begin{minipage}{0.45\textwidth}
    \centering
    \includegraphics[width=\linewidth,trim={0 20mm 0 20mm},clip]{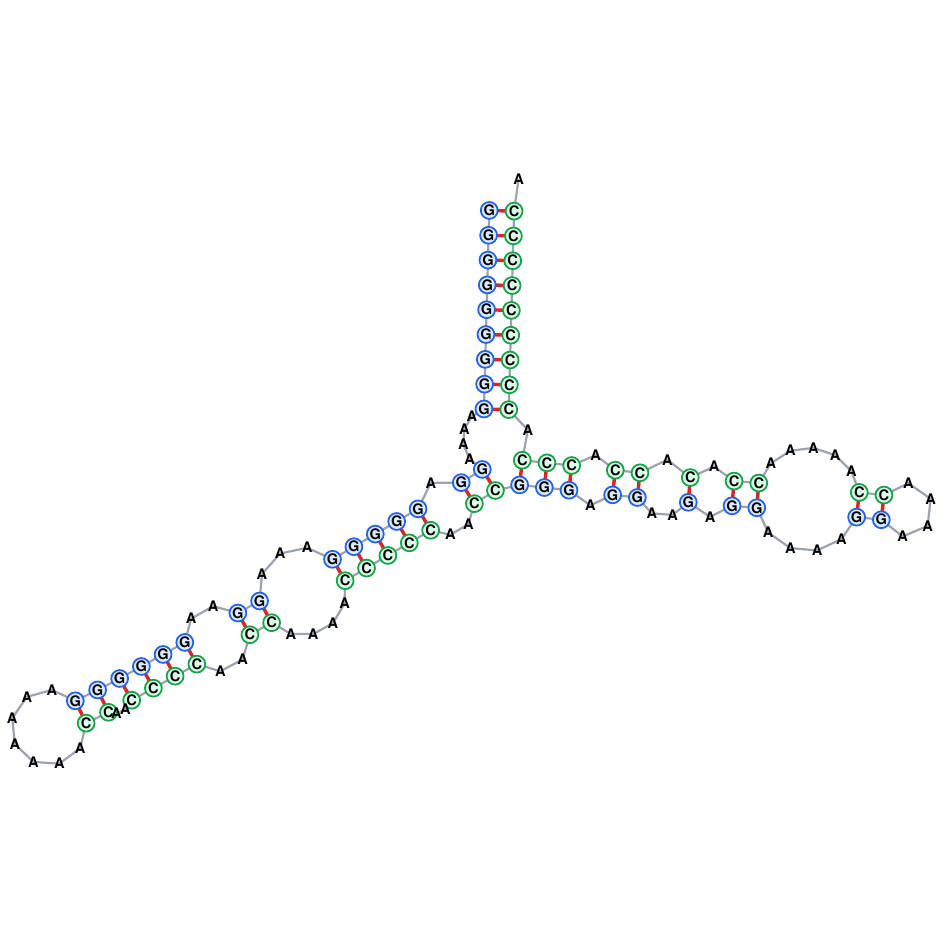}
    \small Rfam27 example
  \end{minipage}
  \hfill
  \begin{minipage}{0.45\textwidth}
    \centering
    \includegraphics[width=\linewidth,trim={0 20mm 0 20mm},clip]{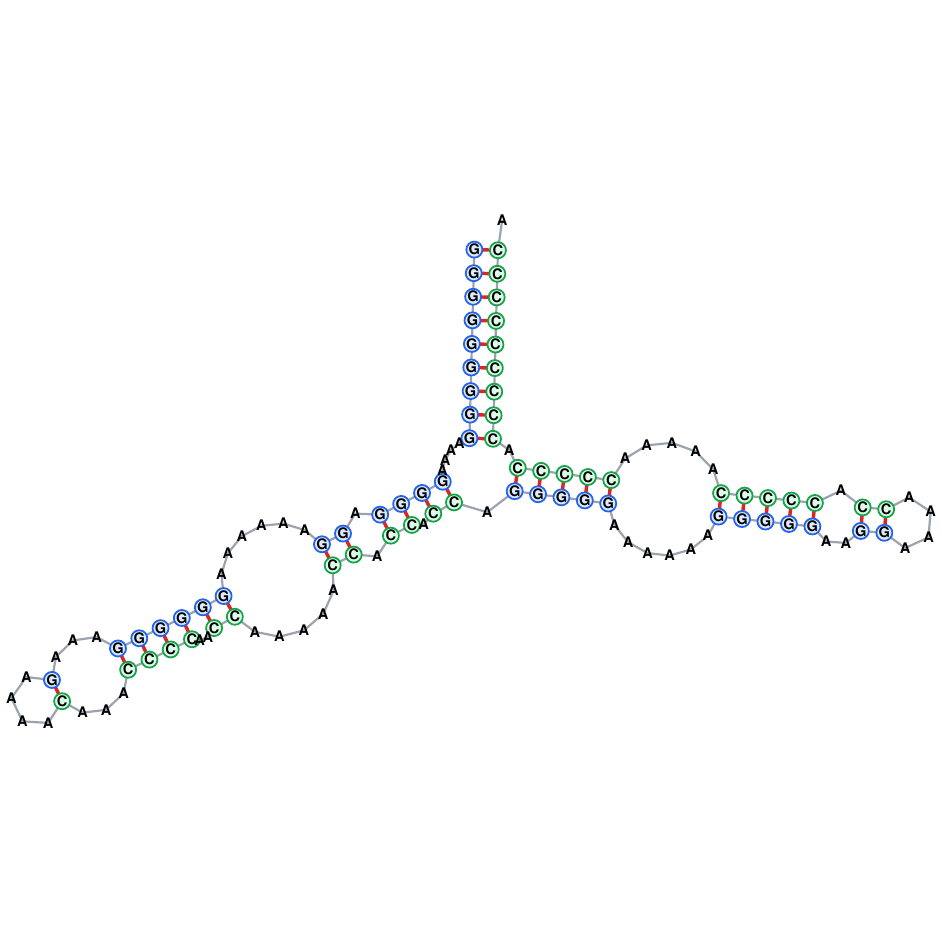}
    \small Nearest RNAsolo100 neighbor
  \end{minipage}
  \caption{Nearest semantic neighbors from disparate dataset chunks under the centered pooled embedding distance. The pair was selected from the pooled-neighbor gallery, not by visual inspection.}
  \label{fig:semantic-neighbor-svgs}
\end{figure}

Figure~\ref{fig:semantic-neighbor-svgs} illustrates the semantic embedding by showing nearest-neighbor structure pairs in disparate datasets.

We perform ridge regression from embeddings to difficulty labels on 2048 random full-structure targets, each evaluated with \(N=100\) samples. The regularization parameter 0.1 was selected via cross-validation. The most stable label was the logit function of best-of-\(N\) NED. Table~\ref{tab:difficulty} shows that GoForth-PSB embeddings transfer better than GoForth-FS embeddings to external structures.

This learned difficulty axis is not model-specific. Across six benchmark chunks and six \(N=100\) methods, difficulty rankings are highly concordant: median pairwise Spearman correlations over method pairs range from \(0.849\) to \(0.946\) for best NED and from \(0.932\) to \(0.969\) for negative log target probability. Appendix~\ref{app:shared-difficulty} gives the per-chunk and leave-one-chunk-out analyses, as well as illustrations of the difficulty fit quality and a quantification of within-dataset and across-dataset similarity induced by the semantic embedding.

\begin{figure}
  \centering
  \begin{minipage}{0.31\textwidth}
    \centering
    \includegraphics[width=\linewidth]{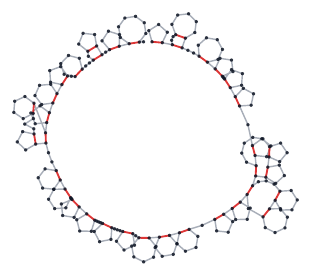}
    \small Target
  \end{minipage}
  \hfill
  \begin{minipage}{0.31\textwidth}
    \centering
    \includegraphics[width=\linewidth]{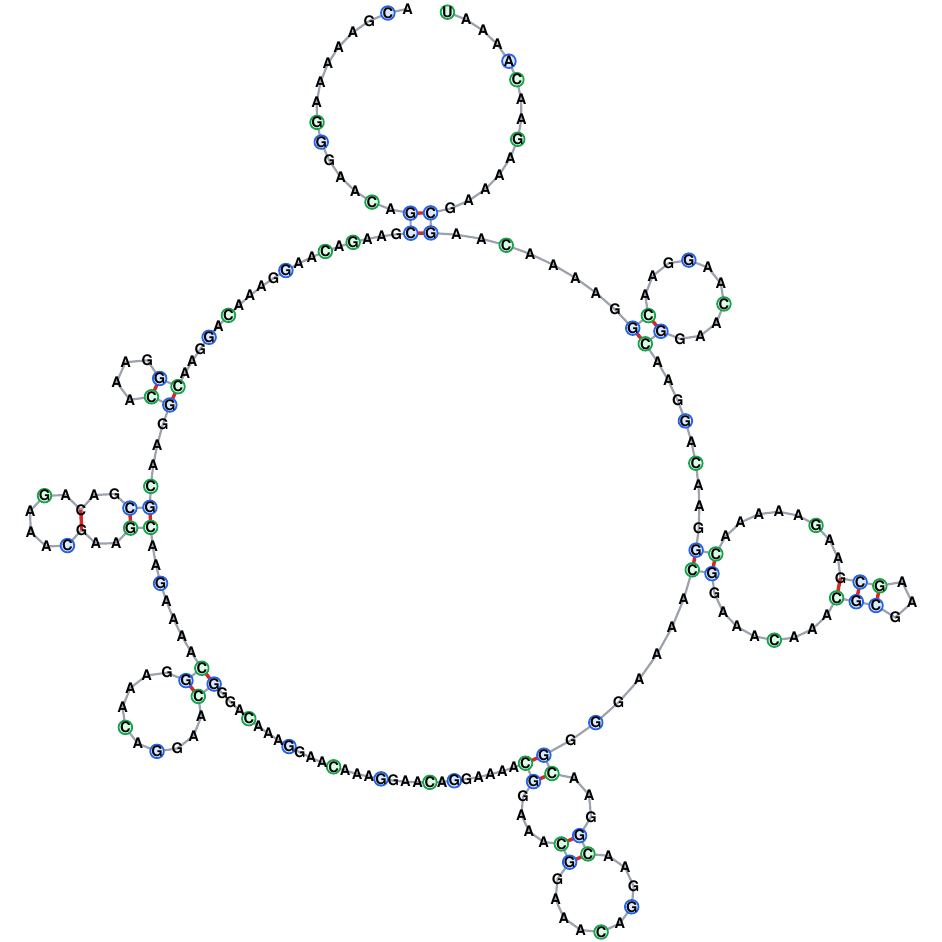}
    \small GoForth MFE
  \end{minipage}
  \hfill
  \begin{minipage}{0.31\textwidth}
    \centering
    \includegraphics[width=\linewidth]{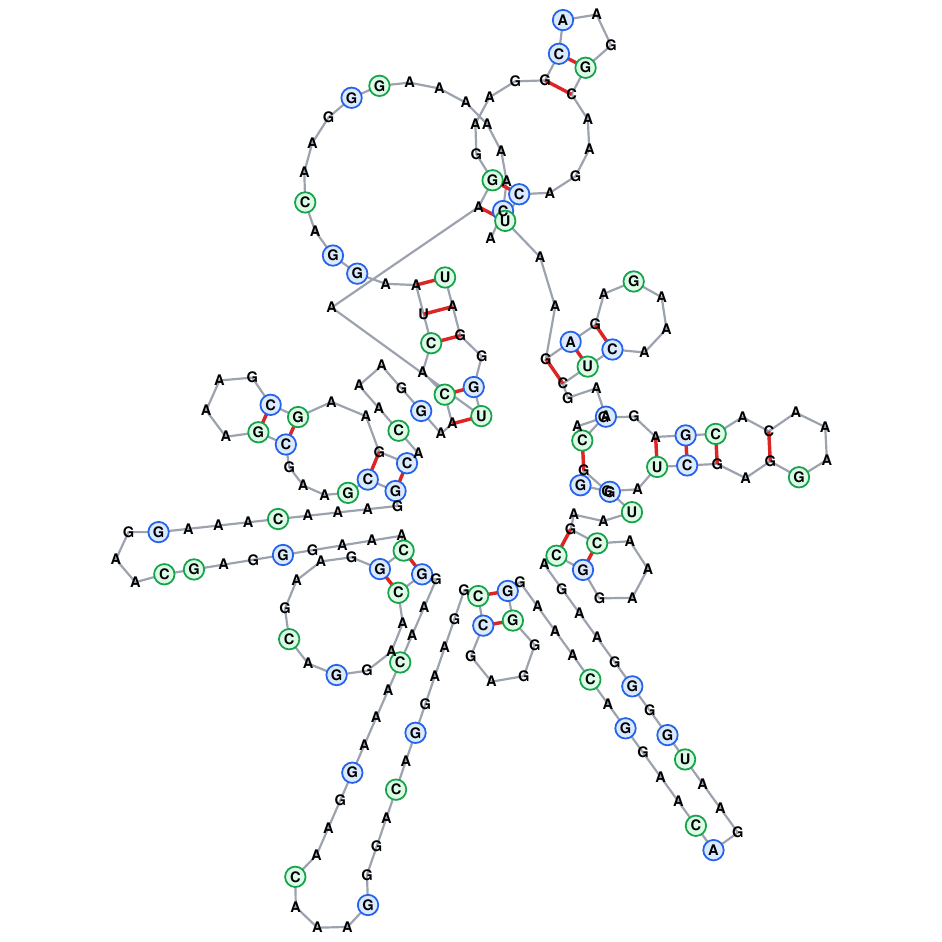}
    \small RNA-Design-LM SL+RL MFE
  \end{minipage}
  \caption{Procedurally generated adversarial structure tuned toward many small loops. The left panel shows the target topology only; black dots mark nucleotide positions and do not denote a realized sequence. The recovered MFEs from GoForth and RNA-Design-LM both miss the target topology, suggesting that adversarial difficulty can transfer across generators.}
  \label{fig:adversarial-svgs}
\end{figure}

Figure~\ref{fig:adversarial-svgs} illustrates a full-structure example that is hard for both GoForth and RNA-Design-LM, which we uncovered with a genetic algorithm that attempts to minimize the designability score while expressing certain target motifs.

\begin{table}[t]
  \centering
  \small
  \caption{Spearman correlation between a linear embedding difficulty score and observed full-structure difficulty, using the transformed best-of-\(N\) NED label. Probes are trained only on the random validation block and evaluated without refitting on each dataset.}
  \label{tab:difficulty}
  \resizebox{\textwidth}{!}{%
  \begin{tabular}{lrrrrrr}
    \toprule
    Embedding model & random train & RNAsolo100 & Rfam27 & unseen & bpRNA strict & Eterna100 \\
    \midrule
    GoForth-FS & 0.560 & 0.451 & 0.513 & 0.388 & 0.519 & 0.540 \\
    GoForth-PSB & \textbf{0.728} & \textbf{0.701} & \textbf{0.637} & \textbf{0.646} & \textbf{0.732} & \textbf{0.699} \\
    \bottomrule
  \end{tabular}}
\end{table}

Additional details of the embedding geometry and designability metric are included in Appendix~\ref{app:shared-difficulty}.

\section{Discussion}

GoForth performs competitively in the domain of full structure design, with excellent computational cost. Since it is trained using only an inexpensive source of data, without an inverse-design teacher, it can be extended easily to accommodate partial design tasks of extreme flexibility in a push-button framework, where it outperforms a state-of-the-art search method and no amortized alternatives exist to our knowledge. We have provided evidence via post-training that temperature-controlled autoregressive sampling interpolates from a posterior sampler to a MAP estimator, connecting our framework with existing objectives for inverse design.

\paragraph{Limitations.}
All current validation is in silico. ViennaRNA is a useful folding oracle, but it is not biology. In particular it is pseudoknot-free. Meanwhile, we only train on RNAs up to \(310\,\nt\). Partial-design benchmarking is also under-standardized.
We present bpRNA and Coding as proposed evaluation scaffolds, not as a settled leaderboard. For external comparisons, possible target or motif overlap across datasets remains an audit item; the learned embedding/designability score gives one possible tool for that audit.

\paragraph{Broader impacts.}
Faster computational RNA design could reduce screening cost in RNA biotechnology, therapeutics, and synthetic biology. The same speed also makes it easier to generate many unvalidated biological hypotheses, so the outputs should be treated with care.

\paragraph{Future work.}
Natural next steps are larger models and length windows, variable-length and ambiguous-base condition languages, biological sequence priors beyond the current uniform prior, richer thermodynamic or experimental folding oracles, and wet-lab validation.

\section{Conclusion}

RNA design is more naturally a conditional sampling problem than a single-solution optimization problem. GoForth operationalizes this view with encoder-decoder autoregressive models for full, partial, base-constrained, and coding-aware secondary-structure design trained from witnessed pairs rather than inverse-design teachers. The trained models define a fast and flexible push-button tool for RNA design within a specified thermodynamic model. Moreover, GoForth defines semantic embeddings of RNA constraints and a practically robust notion of designability. 

\bibliographystyle{plainnat}
\bibliography{references}

\appendix

\section{Architecture and Training Details}
\label{app:model-details}

All GoForth variants use the same encoder-decoder Transformer template and differ only in the condition channels supplied to the encoder. Table~\ref{tab:model-taxonomy} summarizes the model taxonomy.

\begin{table}[h]
  \centering
  \small
  \caption{Model taxonomy. Parameter counts are rounded to the nearest million.}
  \label{tab:model-taxonomy}
  \resizebox{\textwidth}{!}{%
\begin{tabular}{lllll}
    \toprule
    Model & Condition channels & Small params & Large params & Main use \\
    \midrule
    GoForth-FS & full structure & 59M & 200M & full structure design \\
    GoForth-PSB & masked structure + masked bases & 59M & 200M & partial structure/base design \\
    GoForth-PSBC & masked structure + bases + frame/amino acid & 60M & 201M & partial structure/base/coding design \\
    \bottomrule
  \end{tabular}}
\end{table}

\paragraph{Architecture and training.}
The small configuration uses 8 encoder layers, 8 decoder layers, width 512, 8 attention heads, feed-forward width 2048, dropout 0.05, and fractional positional features. The large configuration uses 12 encoder layers, 12 decoder layers, width 768, 12 attention heads, feed-forward width 3072, and dropout 0.05. Supervised training used AdamW with \(\beta=(0.9,0.95)\), weight decay \(0.01\), global gradient clipping at 1.0, packed length-bucketed batches of size 64, bucket width 16, and maximum training length \(310\,\nt\).

\begin{figure}[t]
  \centering
  \includegraphics[width=\textwidth,trim={0 0 0 5mm},clip]{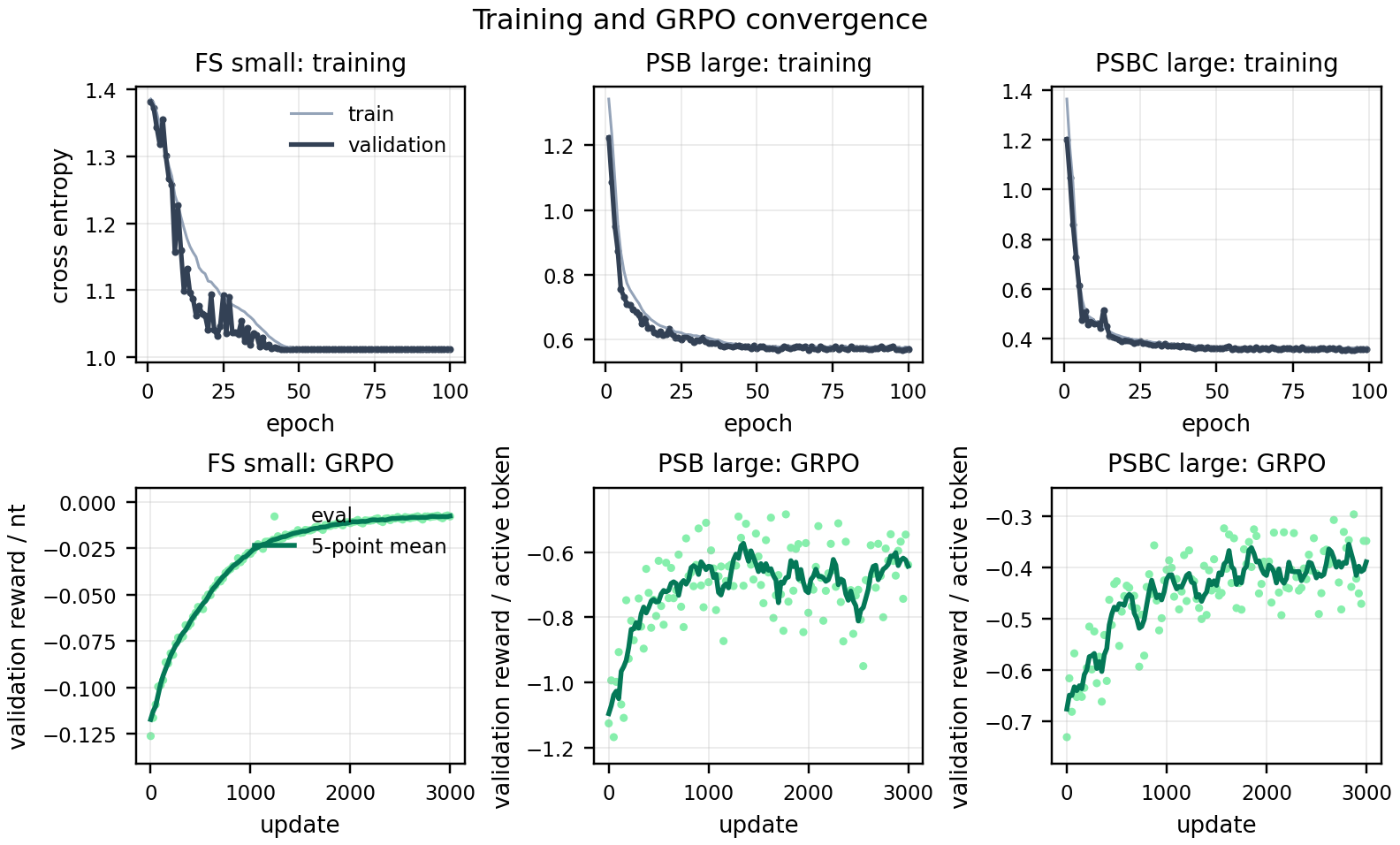}
  \caption{Training and GRPO convergence curves for the principal GoForth-FS, GoForth-PSB, and GoForth-PSBC models. The top row shows training and validation cross-entropy; the bottom row shows GRPO validation reward.}
  \label{fig:training-grpo-convergence}
\end{figure}

The main reservoirs for each training run contain 285{,}039 training and 14{,}961 validation RNA sequences with uniform i.i.d. bases and uniformly random length \(11\)--\(310\,\nt\). The base learning rate is \(3\times10^{-4}\), with 5\% warmup, a held high-learning-rate phase, decay to \(10^{-5}\), a hold phase near \(10^{-5}\), and a final decay to \(10^{-6}\) over 100 epochs. An epoch denotes a single pass over all RNA sequences $x$, but note that for each epoch we sample a new conditionally independent $c$ according to the Vienna model $q(s \,\vert\,  x)$, followed by random masking for the PSB and PSBC models. The masking policy for each model is described in Appendix~\ref{app:benchmark-details}. Protected checkpoints were saved at epochs 50, 75, and 100. Training for each model was performed on a single Nvidia H100 GPU. Robust convergence is validated in Figure~\ref{fig:training-grpo-convergence}.

\paragraph{Post-training.}
GRPO post-training uses target batch size 32, group size 8, learning rate \(3\times10^{-6}\), and \(\lambda_{\mathrm{KL}}=0.05\). The full-structure and mixed-condition GRPO checkpoints in the main tables are update-3000 continuations of the original training runs. Reward evaluation uses hard decoding masks when base, pair, or codon constraints are active, and ViennaRNA scoring was run through a parallel CPU worker pool. Robust convergence is validated in Figure~\ref{fig:training-grpo-convergence}.

\paragraph{Model ablations.}
Two structure-only model ablations were trained during development. These models ignore base and coding constraints, considering $\{ \texttt{L}, \texttt{R}, \texttt{x},  \texttt{?}\}$ and $\{ \texttt{L}, \texttt{R}, \texttt{x}, \texttt{\#},  \texttt{?}\}$ as alternative dictionaries for structure tokens. Relative to the former, the latter allows for the possibility of conditioning on `pairedness' without knowledge of the left/right flavor. These models helped establish that pairedness-only condition tokens were viable, but in fact GoForth-PSB learned to respect base-pair complementarity (i.e., Watson-Crick / wobble pairing) more robustly when supplied full structural constraints, in spite of the apparently richer learning task that it faces in allowing for arbitrary base constraints. We therefore use GoForth-PSB as our general partial-structure/base model and omit the ablations from the principal benchmark tables.

We also trained a GoForth-PSBC variant without the frame token. Training and validation losses were slightly worse, but downstream behavior was qualitatively similar. We keep the frame token because it is cheap, explicit, and achieves modestly better training/validation objective.

\paragraph{Metric conventions.}
For full-structure targets we report best-of-\(N\) MFE hit, unique-MFE hit, Boltzmann target probability, NED, structural distance, and neural generation timing. Benchmark logs also record Vienna scoring time, but the main runtime table excludes it because it depends strongly on CPU worker count and candidate redundancy. Training-length strata are \(\le310\,\nt\) and length-generalization strata are \(>310\,\nt\). For partial conditions, an MFE hit means that at least one unconstrained MFE structure satisfies the active structure tokens while all fixed base/codon constraints are obeyed; masked tokens are ignored. A UMFE hit means that in addition the MFE structure is unique. All base and amino-acid hard-constraint error rates in the reported partial benchmarks were zero.

\section{Dataset Construction Details}
\label{app:benchmark-details}

\paragraph{Synthetic masking policies}
The randomized masking policies that we use to derive training pairs $(x,c)$ from Vienna-witnessed $(x,s)$ are as follows. They are chosen to contain fully arbitrary design constraints within their support, allowing the models to generalize across arbitrary conditional input.

For GoForth-PSB, we uniformly sample independent numbers $N_x$ and $N_s$ of base and structure tokens in $\{0,\ldots L\}$ to mask, where $L$ is the number of nucleotides in the sample $(x,s)$. Then we pick uniformly random subsets of these sizes and mask the corresponding tokens. There are two ways to mask a structure token if it is $\texttt{L}$ or $\texttt{R}$. Thus for each sample, we pick one uniform random variable $p\in [0,1]$ and then for each structure token $\texttt{L}$ or $\texttt{R}$ selected for masking, independently mask it to $\texttt{?}$ with probability $p$, else to $\texttt{\#}$.

For GoForth-PSBC, we additionally sample a uniform divisible-by-3 coding region length $N_\rho$ in $\{0,\ldots L\}$ and, given this, a uniformly random start position $j \in \{0,\ldots L - N_\rho - 1 \}$. Then all positions $i$ receive the frame token $\rho_i = i - j \mod 3$. Moreover, we pick a uniformly random number of codons $N_c \in \{0,\ldots N_\rho / 3\}$ to unmask within the coding region and then pick a uniformly random subset of $N_c$ codons to mask. Specifically we set $\psi_i$ to be the appropriate amino acid token within each of the unmasked codons' subintervals. We keep $\psi_i = \texttt{?}$ everywhere else.

\paragraph{Full structures.}
The full-structure benchmark combines RNAsolo100-derived structures \citep{adamczyk2022rnasolo}, Rfam27 \citep{kalvari2021rfam}, Eterna100 \citep{andersonlee2016principles}, Eterna100-v2 \citep{koodli2021eterna100v2}, and 100 unseen random validation structures. It contains 1031 training-length targets and 60 length-generalization targets.

We also report a separate length-stratified bpRNA-90 strict-canonical set \citep{danaee2018bprna}, with full-structure results reported in Table~\ref{tab:bprna}. This dataset forms the basis of several of our partial structure design tasks.

\begin{table}[t]
  \centering
  \small
  \caption{bpRNA-Annotation benchmark for full structure design, \(N=100\), all targets \(\le310 \, \nt\). Asterisked exact columns omit 6/100 annotations lying outside of ViennaRNA's default MFE/PF loop class, which is not a hard constraint for natural experimental data; NED includes all 100.}
  \label{tab:bprna}
  \begin{tabular}{lrrrr}
    \toprule
    Method & MFE hit* & UMFE hit* & NED & Best \(q\)* \\
    \midrule
    RNA-Design-LM SL & 0.734 & 0.734 & 0.0143 & 0.650 \\
    RNA-Design-LM SL+RL & \textbf{0.766} & \textbf{0.755} & \textbf{0.00836} & \textbf{0.699} \\
    GoForth-FS small $\tau = 0.1$ & 0.713 & 0.713 & 0.0132 & 0.570 \\
    GoForth-FS large $\tau = 0.1$ & 0.702 & 0.691 & 0.0145 & 0.546 \\
    GoForth-PSB small $\tau = 0.1$ & 0.681 & 0.681 & 0.0127 & 0.581 \\
    GoForth-PSB large $\tau = 0.1$ & 0.702 & 0.702 & 0.0125 & 0.575 \\
    GoForth-PSBC small $\tau = 0.1$ & 0.691 & 0.691 & 0.0126 & 0.584 \\
    GoForth-PSBC large $\tau = 0.1$ & 0.713 & 0.713 & 0.0122 & 0.588 \\
    \bottomrule
  \end{tabular}
\end{table}


\paragraph{Witnessed partial structure design tasks.}
The bpRNA dataset consists of sequence-structure pairs $(x,s)$ that annotate sequences with structures. To obtain Vienna-witnessed data, we replace each $s$ with the MFE fold of $x$ and then apply some masking to $(x,s)$. However, we can also view the original pair $(x,s)$, corrupted by some masking, as being witnessed by natural experiment. We call the resulting dataset bpRNA-Annotation.

To produce the Coding dataset, we curate a list of 100 short peptides less than 100 amino acids in length, then apply the algorithm of \citep{fornace2026direct} to sample a sequence $x$ that admits stable folding. Then we let $s$ be its Vienna MFE structure, which we assume to be robust enough to be designable. Then we apply some masking to the pair $(x,s)$.

\paragraph{Motif-based masking policies}


Policy 1 reveals one to three terminal hairpins from bpRNA annotations. Policy 2 additionally fixes loop bases and a pairedness/accessibility halo. Policy 3 reveals a codon-aligned local structure window while preserving the full peptide. The bpRNA policies have 76 eligible targets after terminal-hairpin filtering; the coding-window policy has 100 targets.

\section{Length Extrapolation}\label{app:length}

\begin{table}[t]
  \centering
  \small
  \caption{Length-generalization slice (\(>310\,\nt\)) of the full-structure benchmark. These 60 targets are outside our training length range but within the range of RNA-Design-LM.}
  \label{tab:length-generalization}
  \begin{tabular}{lrrrr}
    \toprule
    Method & MFE hit & NED & Best \(q(\struct\given\seq)\) & Structural distance \\
    \midrule
    RNA-Design-LM SL & 0.283 & 0.158 & 0.196 & 47.8 \\
    RNA-Design-LM SL+RL & \textbf{0.433} & \textbf{0.058} & \textbf{0.336} & \textbf{13.6} \\
    GoForth-FS small $\tau = 0.1$ & 0.317 & 0.077 & 0.116 & 17.5 \\
    GoForth-FS large $\tau = 0.1$ & 0.383 & 0.064 & 0.132 & 13.7 \\
    GoForth-PSB small $\tau = 0.1$ & 0.333 & 0.073 & 0.130 & 18.4 \\
    GoForth-PSB large $\tau = 0.1$ & 0.317 & 0.075 & 0.126 & 17.9 \\
    GoForth-PSBC small $\tau = 0.1$ & 0.300 & 0.086 & 0.123 & 22.0 \\
    GoForth-PSBC large $\tau = 0.1$ & 0.317 & 0.073 & 0.150 & 19.2 \\
    \bottomrule
  \end{tabular}
\end{table}

\begin{figure}
  \centering
  \begin{minipage}{0.49\textwidth}
    \centering
    \includegraphics[width=\linewidth]{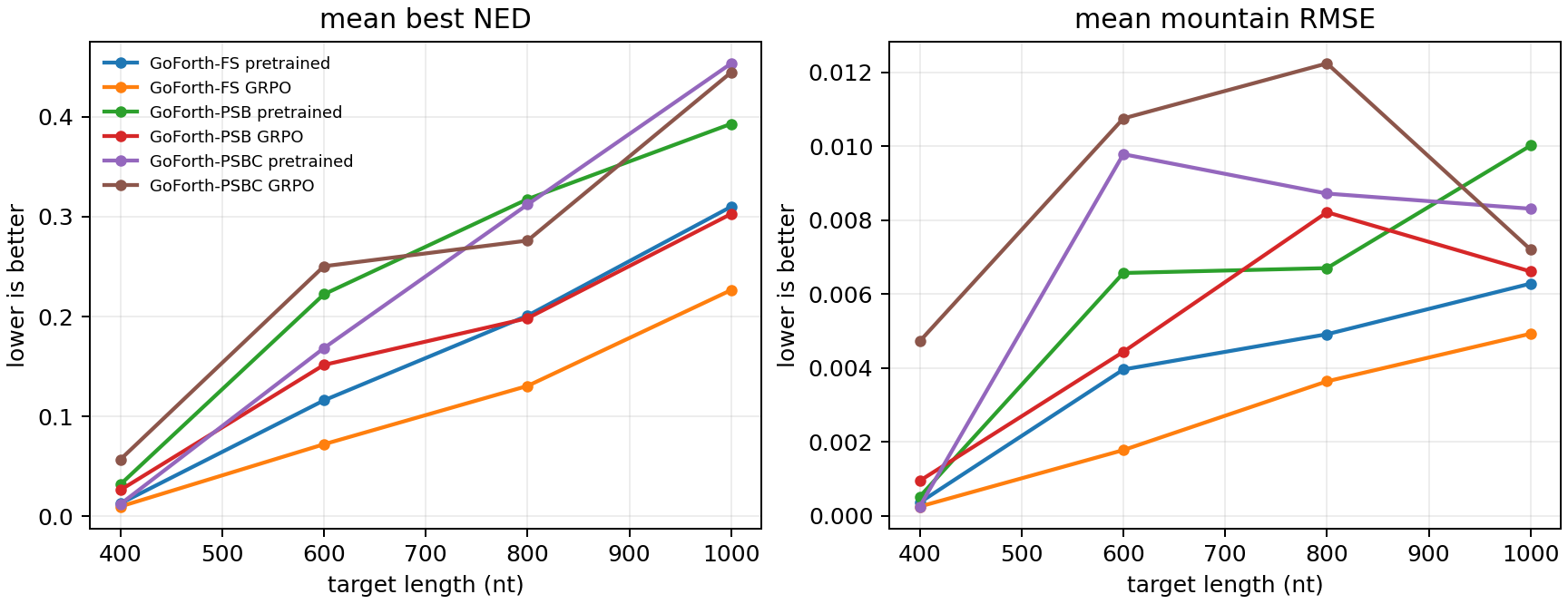}
  \end{minipage}
  \hfill
  \begin{minipage}{0.49\textwidth}
    \centering
    \includegraphics[width=\linewidth]{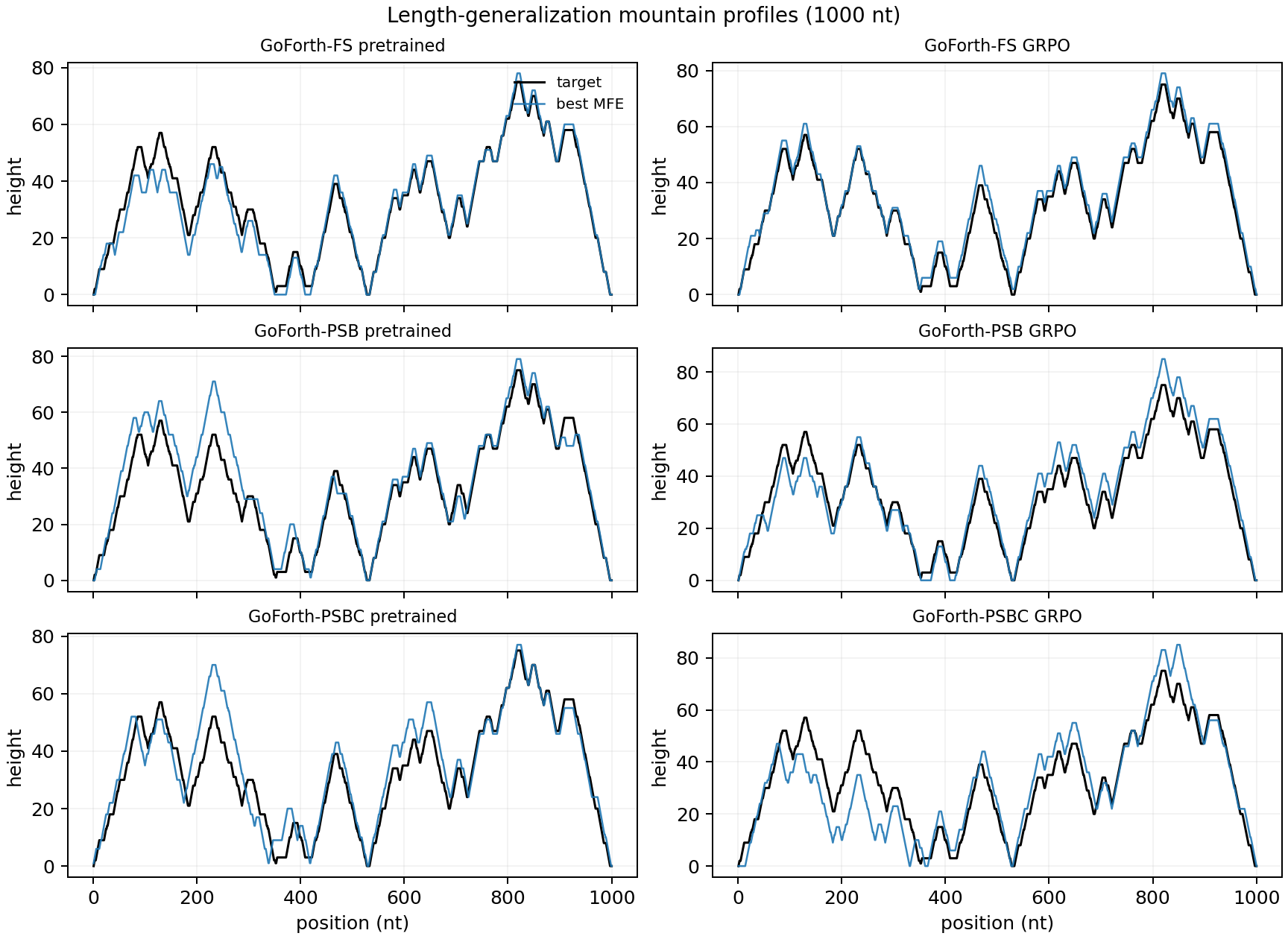}
  \end{minipage}
  \caption{Illustrative length extrapolation diagnostics. Left: best-of-\(N\) NED and mountain-profile RMSE for random Vienna-witness targets of length 400--1000. Right: mountain-profile comparison for a longest-target example.}
  \label{fig:length-demo}
\end{figure}

\begin{figure}
  \centering
  \begin{minipage}{0.45\textwidth}
    \centering
    \includegraphics[width=\linewidth]{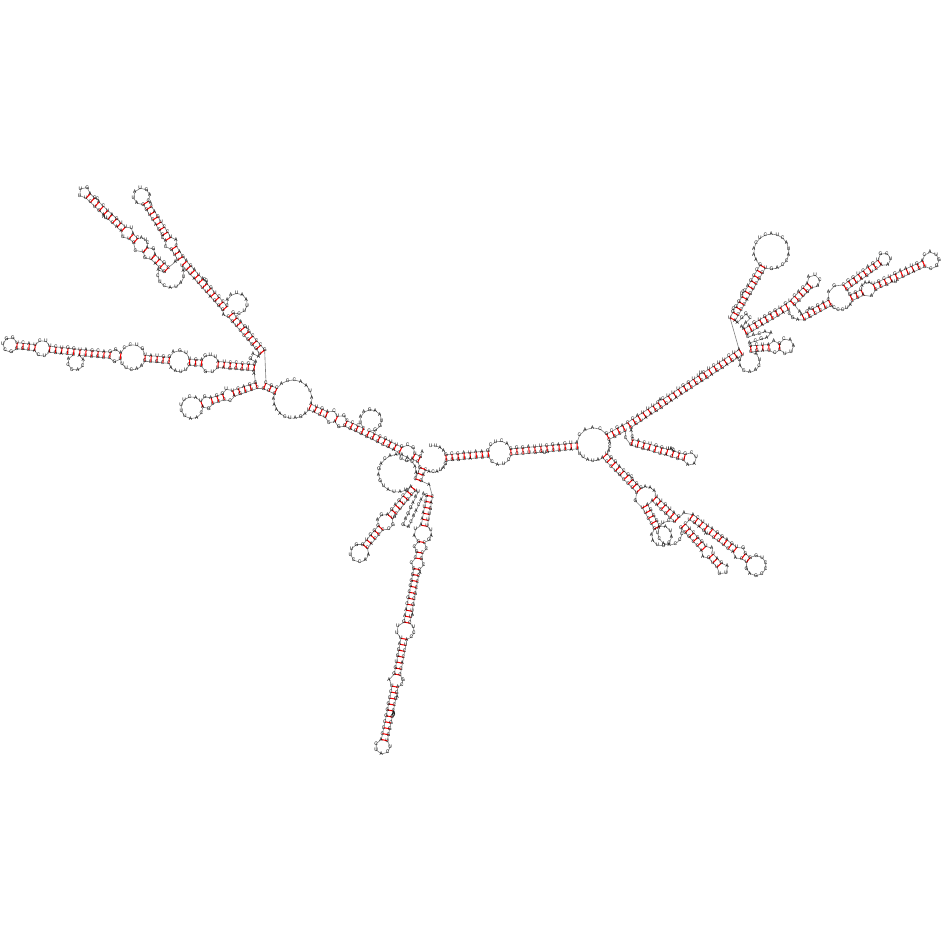}
    \small \(L=1000\) target
  \end{minipage}
  \hfill
  \begin{minipage}{0.45\textwidth}
    \centering
    \includegraphics[width=\linewidth]{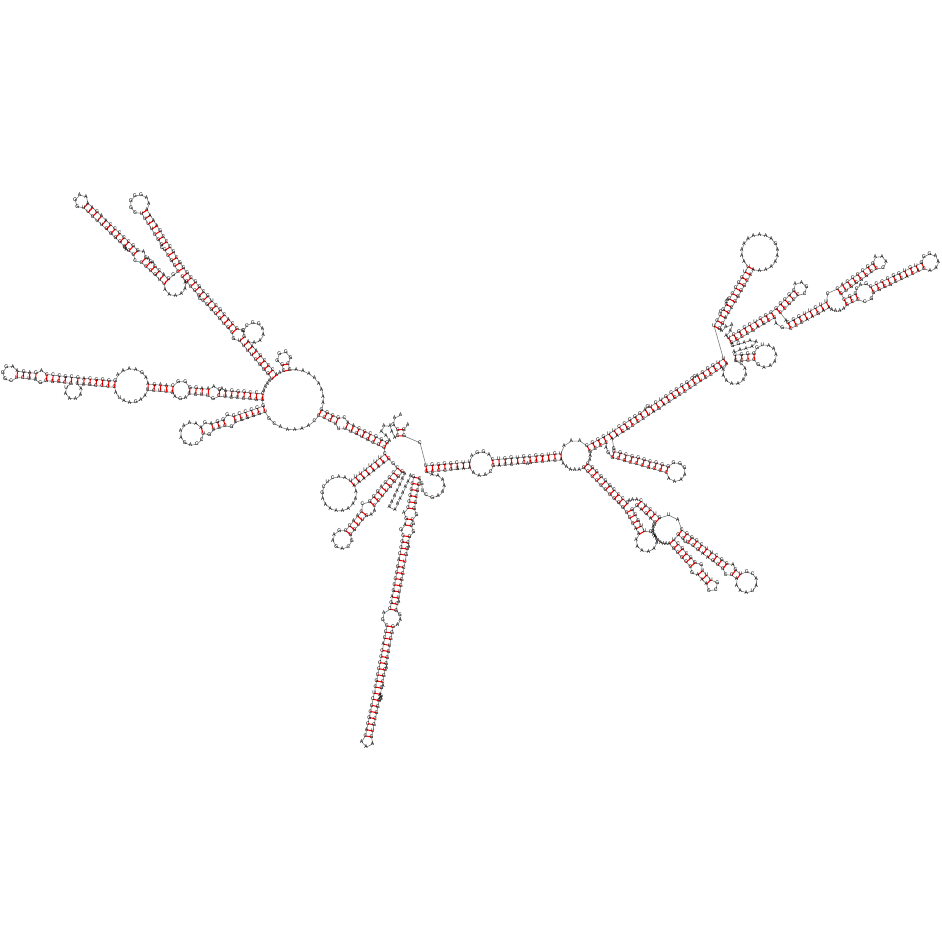}
    \small GoForth-FS GRPO MFE
  \end{minipage}
  \caption{Representative long-chain target and recovered MFE structure. Exact pair recovery is poor at this length, but the mountain profile (cf. Figure~\ref{fig:length-demo}) preserves partial global similarity.}
  \label{fig:length-svgs}
\end{figure}

\begin{figure}
  \centering
  \includegraphics[width=0.72\textwidth]{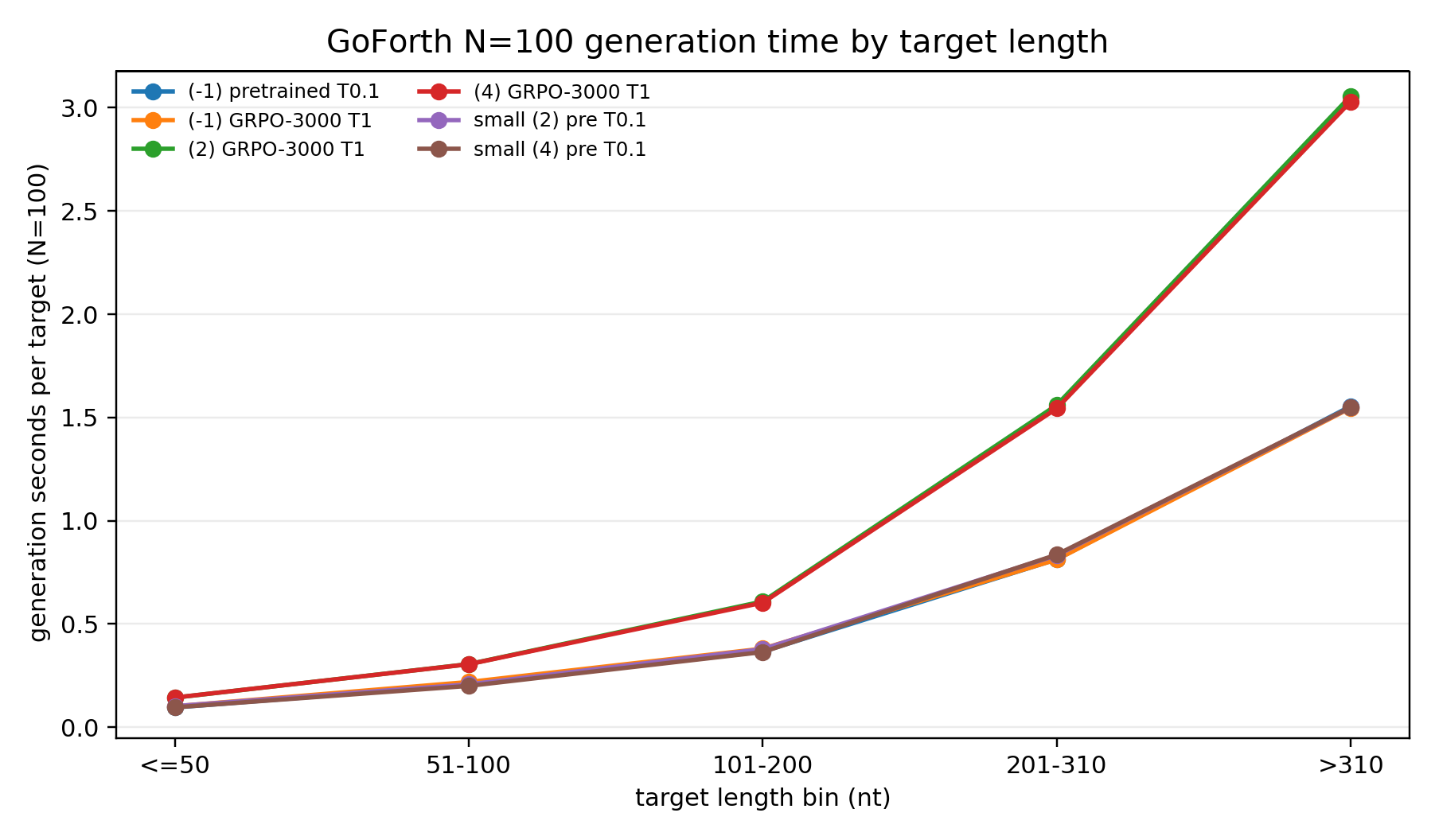}
  \caption{Best-of-\(100\) neural generation time as a function of target length on the full-structure benchmark. Points are mean per-target generation times within length bins.}
  \label{fig:runtime-length}
\end{figure}

All methods degrade outside of our training length range of up to $310 \, \nt$, though RNA-Design-LM was trained on sequences up to $500 \, \nt$, so we keep this data slice in Table~\ref{tab:length-generalization} separate from the headline benchmark. Figures~\ref{fig:length-demo} and~\ref{fig:length-svgs} show illustrative length extrapolation experiments. We observe that coarse mountain-profile statistics retain similarity even for long RNAs.
Figure~\ref{fig:runtime-length} shows how neural generation time scales with target length.

\section{Embedding and Designability}
\label{app:shared-difficulty}

In Section \ref{sec:embedding} we introduce embeddings to predict designability from a condition $c$. A complementary question is whether independently trained generators agree on which full-structure targets are hard. We analyzed \(N=100\) best-of-\(N\) summaries for six GoForth variants, including pretrained and post-trained checkpoints not all shown in the main tables. For each dataset chunk, we computed Spearman correlations over target difficulty rankings across all 15 method pairs, using both best-of-\(N\) NED and negative log target probability.

\begin{table}[h]
  \centering
  \small
  \caption{Within-chunk cross-method agreement in full-structure target difficulty. Entries give median/min/max Spearman correlation across the 15 method pairs.}
  \label{tab:shared-difficulty-within}
  \begin{tabular}{lrrr}
    \toprule
    Chunk & Targets & Best NED & \(-\log q(\struct\given\seq)\) \\
    \midrule
    bpRNA strict & 100 & 0.943 / 0.909 / 0.974 & 0.949 / 0.881 / 0.984 \\
    unseen & 100 & 0.928 / 0.865 / 0.975 & 0.969 / 0.928 / 0.990 \\
    Eterna100 & 100 & 0.939 / 0.892 / 0.975 & 0.959 / 0.917 / 0.985 \\
    Eterna100-v2 & 100 & 0.933 / 0.893 / 0.977 & 0.950 / 0.907 / 0.980 \\
    Rfam27 & 27 & 0.849 / 0.764 / 0.936 & 0.932 / 0.878 / 0.969 \\
    RNAsolo100 & 764 & 0.946 / 0.897 / 0.982 & 0.940 / 0.915 / 0.987 \\
    \bottomrule
  \end{tabular}
\end{table}

We also performed a leave-one-chunk-out analysis. For each held-out chunk, method reliability weights were estimated from all other chunks after chunk-local standardization, then applied to the held-out targets. The resulting consensus ranking was compared against the individual held-out method rankings. This tests whether the difficulty axis is tied to one dataset family or remains stable when the dataset is omitted from the consensus definition.

\begin{table}[h]
  \centering
  \small
  \caption{Leave-one-chunk-out consensus agreement. The first two metric columns report median/min Spearman correlation between the out-of-chunk consensus and the held-out method rankings. The final column compares the two out-of-chunk consensus rankings on the held-out targets.}
  \label{tab:shared-difficulty-loco}
  \begin{tabular}{lrrr}
    \toprule
    Held-out chunk & LOCO best NED & LOCO \(-\log q\) & NED vs. \(-\log q\) \\
    \midrule
    bpRNA strict & 0.967 / 0.960 & 0.984 / 0.936 & 0.773 \\
    unseen & 0.970 / 0.925 & 0.990 / 0.955 & 0.897 \\
    Eterna100 & 0.973 / 0.952 & 0.980 / 0.959 & 0.885 \\
    Eterna100-v2 & 0.971 / 0.946 & 0.973 / 0.956 & 0.880 \\
    Rfam27 & 0.917 / 0.858 & 0.969 / 0.918 & 0.933 \\
    RNAsolo100 & 0.972 / 0.955 & 0.974 / 0.961 & 0.907 \\
    \bottomrule
  \end{tabular}
\end{table}

The chunks are disjoint target sets. The LOCO analysis therefore estimates the consensus definition out of chunk and then measures held-out agreement; it is not a same-target train/test prediction experiment. Within this limitation, the high correlations support the working interpretation that designability is a shared empirical property, not merely a failure mode of one generator.

\begin{figure}
  \centering
  \includegraphics[width=0.8\textwidth]{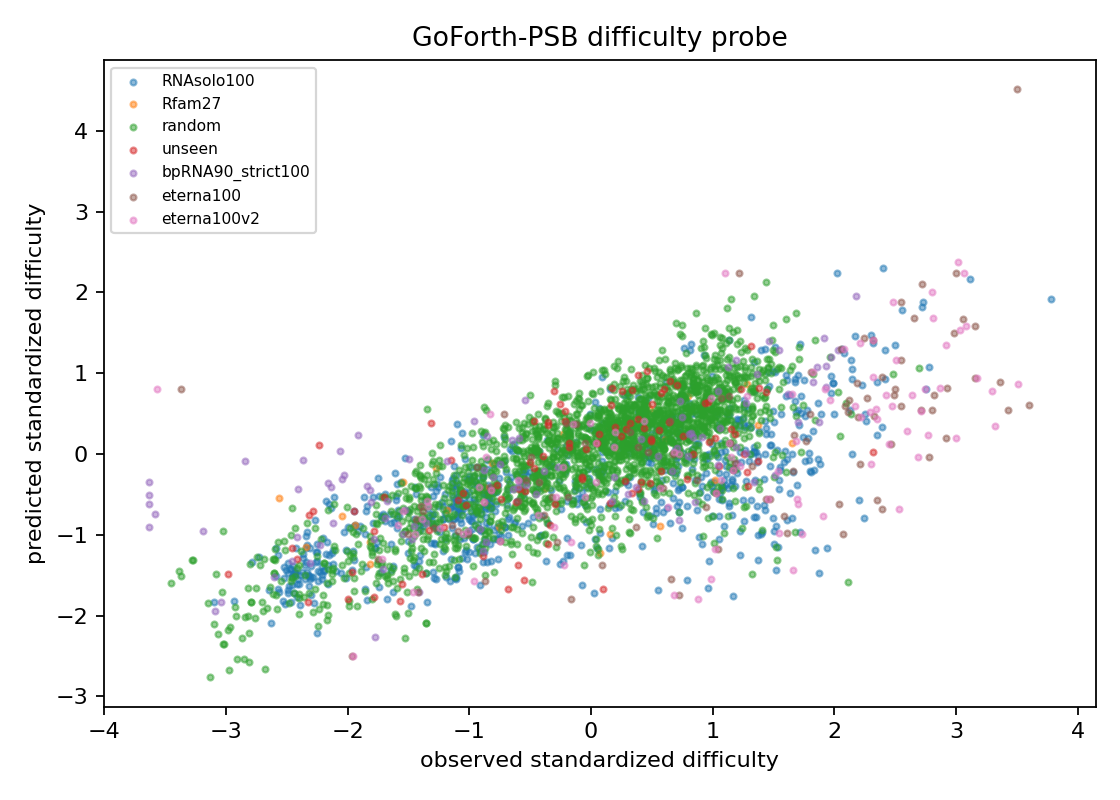}
  \caption{Predicted versus observed full-structure difficulty for GoForth-PSB, using centered and normalized mean-pooled encoder embeddings.}
  \label{fig:model2-difficulty}
\end{figure}

Figure~\ref{fig:model2-difficulty} shows the full-structure difficulty probe.

\begin{figure}
  \centering
  \includegraphics[width=0.8\textwidth]{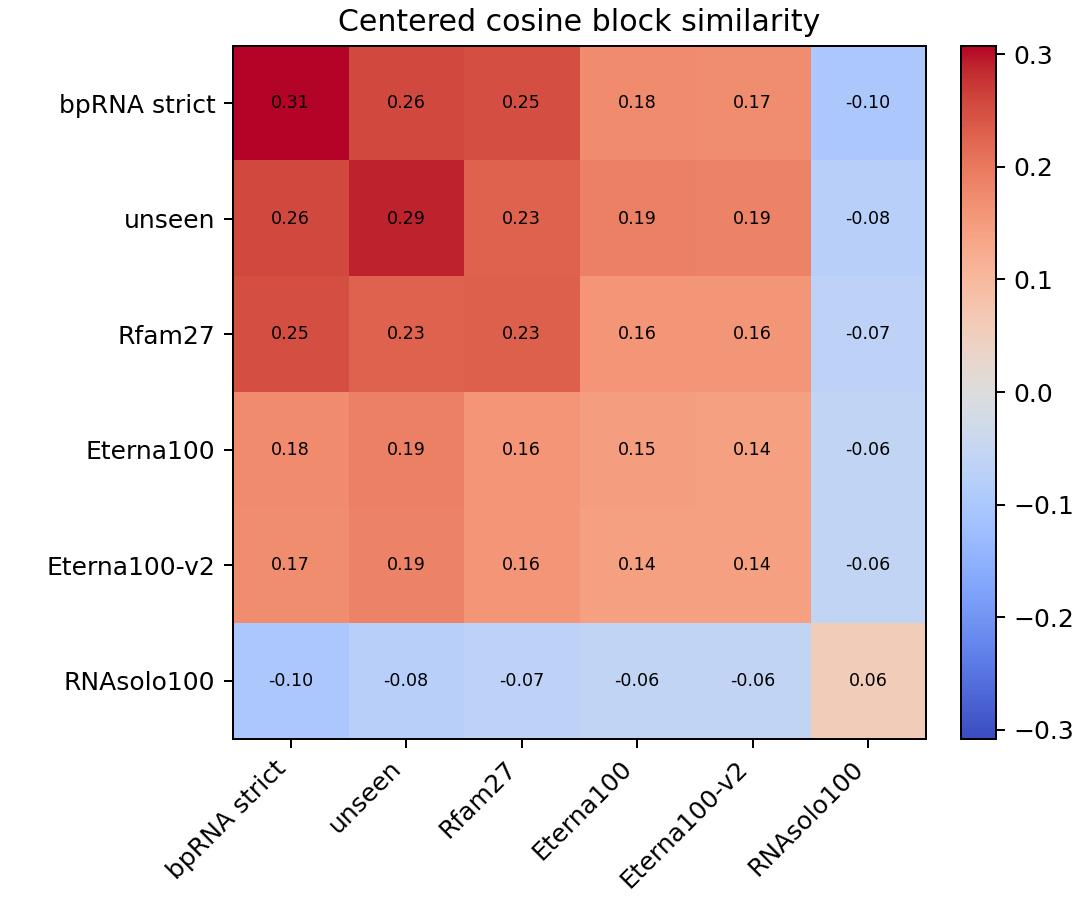}
  \caption{Dataset-level similarity from centered, normalized GoForth condition embeddings, ordered from most to least self-similar on the diagonal. Each entry is the inner product between dataset mean embeddings, yielding a positive-semidefinite block summary.}
  \label{fig:dataset-similarity}
\end{figure}

Figure~\ref{fig:dataset-similarity} shows the dataset-block similarity matrix induced by our embedding.

\section{Additional Experiment Notes}
\label{app:additional-figures}

\paragraph{SMC and backtracking.}
We tested sequential Monte Carlo (SMC) sampling and a backtracking/filtering approach on small smoke benchmarks. Note that SMC can be used to sample autoregressively from an annealed distribution in a more principled fashion by retaining particle weights. However, these alternatives did not clearly dominate ordinary temperature-controlled autoregressive sampling at the chosen temperatures. They are therefore optional search refinements rather than the default algorithm in the current draft.

\paragraph{Hit-based rewards.}
We also tried GRPO rewards based directly on MFE-hit and unique-MFE-hit indicators, analogous in spirit to terms used in RNA-Design-LM. These rewards did not substantially change the full or partial benchmark conclusions, possibly because we do not consider the same curated dataset for RL. This is a useful diagnostic: for rewards tied to \(q(\condition\given\seq)\), GRPO mostly sharpens an existing conditional proposal, so limited gains suggest that the pretrained sampler with low-temperature is already near the relevant oracle-annealed regime.




\paragraph{Additional Model Benchmarks}
\label{app:model-matrix-extra}

The main text includes the large-model pretrained rows most relevant for the paper narrative. For completeness, Table~\ref{tab:small-partial-extra} records the small-model pretrained GoForth-PSB/PSBC rows from the same partial-design benchmark family. These rows are useful for auditing scale effects.

\begin{table}[h]
  \centering
  \small
  \caption{Additional small-pretrained partial-design rows at \(N=100\), \(\tau=0.1\).}
  \label{tab:small-partial-extra}
  \resizebox{\textwidth}{!}{%
  \begin{tabular}{llrrrr}
    \toprule
    Benchmark & Model & Targets & Hit@100 & UMFE@100 & Best MFE error \\
    \midrule
    bpRNA-Masked & GoForth-PSB small pretrained & 100 & 0.530 & 0.450 & 0.022 \\
    bpRNA-Annotation & GoForth-PSB small pretrained & 100 & 0.300 & 0.290 & 0.100 \\
    bpRNA-Masked & GoForth-PSBC small pretrained & 100 & 0.650 & 0.510 & 0.015 \\
    bpRNA-Annotation & GoForth-PSBC small pretrained & 100 & 0.250 & 0.210 & 0.131 \\
    Coding-Masked & GoForth-PSBC small pretrained & 100 & 0.890 & 0.670 & 0.002 \\
    P1 bpRNA hairpins & GoForth-PSB small pretrained & 76 & 1.000 & 1.000 & 0.000 \\
    P2 bpRNA loop+halo & GoForth-PSB small pretrained & 76 & 0.737 & 0.737 & 0.018 \\
    P1 bpRNA hairpins & GoForth-PSBC small pretrained & 76 & 1.000 & 1.000 & 0.000 \\
    P2 bpRNA loop+halo & GoForth-PSBC small pretrained & 76 & 0.737 & 0.724 & 0.019 \\
    P3 Coding-window & GoForth-PSBC small pretrained & 100 & 0.870 & 0.780 & 0.007 \\
    \bottomrule
  \end{tabular}}
\end{table}

\end{document}